\documentclass[prl,twocolumn,superscriptaddress,showpacs,floatfix]{revtex4-1}
\usepackage{amsfonts}
\usepackage{amsmath}
\usepackage{amssymb}
\usepackage{graphicx}

\begin{document}

\title{Entanglement of formation for a class of $(2\otimes d)$-dimensional systems}
\date{\today}

\author{F. Lastra}
\affiliation{Departamento de F\'{\i}sica, Universidad de Santiago de
Chile, USACH, Casilla 307 Correo 2 Santiago, Chile}
\affiliation{Center for the Development of Nanoscience and Nanotechnology, 9170124, Estaci\'on Central, Santiago, Chile} 

\author{C. E. L\'opez}
\affiliation{Departamento de F\'{\i}sica, Universidad de Santiago de
Chile, USACH, Casilla 307 Correo 2 Santiago, Chile}
\affiliation{Center for the Development of Nanoscience and Nanotechnology, 9170124, Estaci\'on Central, Santiago, Chile}

\author{L. Roa}
\affiliation{Departamento de F\'{\i}sica, Universidad de Concepci\'on, Casilla 160-C, Concepci\'on, Chile.}

\author{J. C. Retamal}
\affiliation{Departamento de F\'{\i}sica, Universidad de Santiago de
Chile, USACH, Casilla 307 Correo 2 Santiago, Chile}
\affiliation{Center for the Development of Nanoscience and Nanotechnology, 9170124, Estaci\'on Central, Santiago, Chile}

\begin{abstract}
Currently the entanglement of formation can be calculated analytically for mixed states in a $(2\otimes2)$-dimensional Hilbert space. For states in higher dimensional Hilbert space a closed formula for quantifying entanglement does not exist. In this regard only entanglement bounds has been found for estimating it.
In this work, we find an analytical expression for evaluating the entanglement of formation for bipartite ($2\otimes d$)-dimensional mixed states. 

\end{abstract}
\pacs{03.67 Mn, 03.65 Yz}

\maketitle

Entanglement in multipartite quantum systems is an issue of fundamental interest in the fields of Quantum Information and Quantum Computation \cite{horo1,Geza,wal,alm,lau}. Although we know  the entanglement  of formation of a general mixed state of two qubits \cite{Wootters},  for general bipartite mixed states in higher dimensional Hilbert space is unknown.  A particular result is known for entanglement of formation of isotropic states in arbitrary dimensions \cite{ter,rungta}. For general states the efforts have led to bounds for estimating the amount of entanglement \cite{Ger,Chen05,Gunhe}. 

In this work we find an analytical expression for evaluating the entanglement of formation for a class of $2\otimes d$ dimensional mixed states.  Specifically we consider the calculation of EoF  for  $2\otimes d$ bipartite mixed  states obtained from tripartite $2\otimes 2\otimes d$ pure states after tracing out one of the  two-dimensional subsystem. Two physical models  that meet such requirements are studied and the EoF is compared  with a lower bound for EoF in higher dimensional systems \cite{Chen05}. 

\begin{figure}[hb]
\includegraphics[width=65mm]{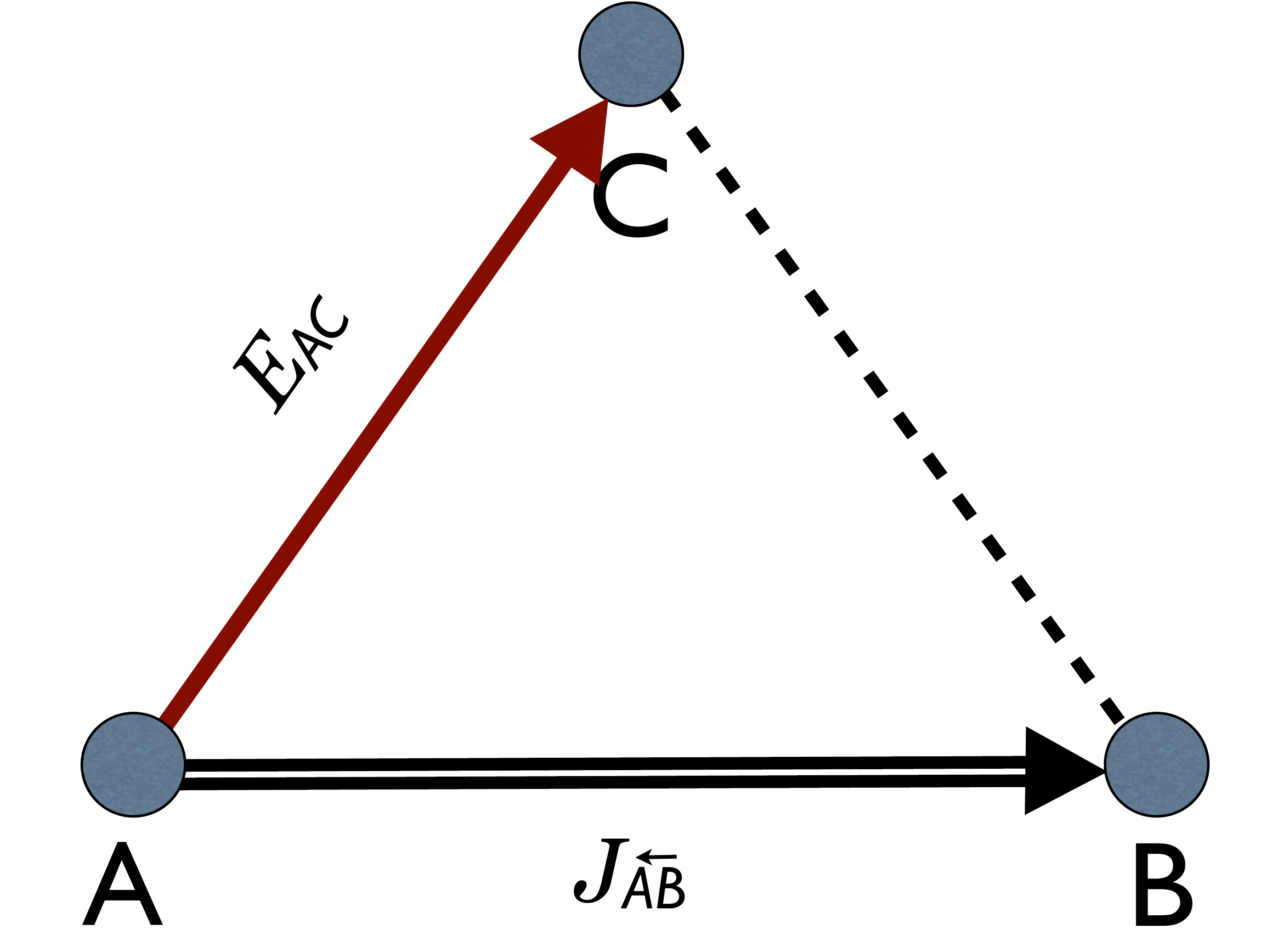}	
\caption{Scheme for the Koashi and Winter identity. Double line represents the classical correlation where an optimal measurement is carried out on system $B$.
Single arrow line means entanglement of formation of the bipartite state $\rho_{AC}$.} \label{scheme}
\end{figure}

Our treatment appeals to  recent developments in the study of quantum correlations for bipartite quantum systems. The extension of  classical mutual information to its quantum counterpart have shown that quantum correlations go beyond entanglement \cite{gro}.  The  difference between quantum mutual information and classical correlation is defined as the quantum discord \cite{Ollivier}, which measures the quantumness or Bayesian degree of a bipartite state \cite{Modi}. A closed formula for  quantum discord in a class of two-qubit states was achieved by Luo \emph{et al.}~\cite{Luo} and by Ali \emph{et al.}~\cite{xstate}. The study of such correlations has attracted much attention in last years \cite{Caves,Almeida,Lidar,Maniscalco,Aolita,caldeira}. At this respect, Koashi and Winter found an identity for a tripartite ($ABC$) pure quantum state \cite{koashi}, that relates the entanglement of formation (EoF) $E_{AC}$ of the reduced pair $AC$ to the classical correlation (CC) $J^{\leftarrow}_{AB}$ \cite{hen,ved}  in the pair $AB$ and to the von Neumann entropy of the reduced state of $A$. Thus, the  identity maps the problem of finding the EoF between $A$ and $C$ to the problem of finding the CC of the partition $AB$. This is schematically represented in Fig. \ref{scheme}.
It is important to emphasize that the identity is not imposing restrictions on the dimensionality of the involved subsystems but on the purity of the tripartite state \cite{koashi}.

Let us consider a tripartite pure state of systems $A$, $B$, and $C$,  where $A$ and $B$ are qubits and $C$ is a qudit ($d$-dimensional system).
The correlations are distributed among the subsystems~\cite{koashi} as follows:
\begin{equation}
E_{AC}+J^{\leftarrow}_{AB}=S_{A}, \label{eof}
\end{equation}
where $S_{A}$ is the von Neumann entropy for the reduced state $\rho_A$. The CC \cite{hen,ved} is defined by the optimization:
\begin{equation}
J^{\leftarrow}_{AB}=\max_{\{\Pi_{x}\}}\left[S_{A}-S(\rho_{AB}|\{\Pi_{x}\})\right],
\label{J}
\end{equation}
$S(\rho_{AB}|\{\Pi_{x}\})=\sum_{x}p_{x}S_{A}^{x}$ is the quantum conditional entropy of $A$ averaged on all possible outcomes of the measurement $\{\Pi_{x}\}$ performed on the subsystem $B$,
$p_{x}=\text{Tr}\{\Pi_{x}\rho_{AB}\Pi_{x}\}$ is the probability with outcome $x$, and
$\rho_{A}^{x}=\text{Tr}_{C}\{\Pi_{x}\rho_{AB}\Pi_{x}\}/p_{x}$.
Because of the nonsymmetric nature of the CC and by choosing three different vertices in Fig. \ref{scheme} we find that there are six relations of the kind (\ref{eof}).
Denoting the ordering $ABC$ for Fig.~\ref{scheme}, by even permutations we can obtain two additional equalities for ordering $CAB$ and $BCA$.
On the other hand, by exchanging $B$ and $C$ on Fig.~\ref{scheme}  obtain an equality for the ordering $ACB$: $E_{AB}+J^{\leftarrow}_{AC}=S_{A}$, which gives rise to equalities for $BAC$ and $CBA$ after permutations. Thus, according with equations (\ref{eof}) and (\ref{J}) the EoF $E_{AC}$ is given by:
\begin{equation}
E_{AC}=\min_{\Pi_{x}}\left[\emph{S}(\rho_{AB}|\{\Pi_{x}\})\right].
\label{condentro}
\end{equation}
Let us assume a general $2\otimes2\otimes d$ tripartite pure state given by:
\begin{eqnarray}
|\Psi\rangle_{ABC}&=&c_{gg}|gg\rangle_{AB}|\psi_1\rangle+c_{ge}|ge\rangle_{AB}|\psi_2\rangle  \nonumber \\
&+& c_{eg}|eg\rangle_{AB}|\psi_3\rangle+c_{ee}|ee\rangle_{AB}|\psi_4\rangle,
\label{gstate}
\end{eqnarray}
where $|\psi_i\rangle$ are arbitrary states of the $d$-dimensional system $C$.
By tracing out one qubit we are led with a $(2\otimes d)$-dimensional mixed state.
The relation (\ref{condentro}) allows us to obtain the entanglement of formation for that bipartite mixed states,
inasmuch as we can solve the optimization process to obtain the conditional entropy associated with the $2\otimes 2$ partition.
The problem of optimizing  $\emph{S}(\rho_{AB}|\{\Pi_{x}\}$ can be solved at least numerically.
However, there are some special classes of states (\ref{gstate}) where the $2\otimes2$ partition corresponds to a $X$ state. This is the case for
 $\langle\psi_1|\psi_2\rangle=\langle\psi_1|\psi_3\rangle=0$ and $\langle\psi_4 | \psi_2\rangle=\langle\psi_4 |\psi_3\rangle=0$.
For such classes an analytical recipe is available for solving the optimization process required to obtain $E_{AC}$~\cite{Luo,xstate}.
When these conditions hold, in the basis $\{|gg\rangle\equiv |0\rangle,|ge\rangle\equiv |1\rangle,|eg\rangle\equiv |2\rangle,|ee\rangle\equiv |3\rangle\}$ we lead with a state of the form:
\begin{equation}
\text{Tr}_C\left(|\Psi\rangle_{ABC}\langle\Psi|_{ABC}\right)=\left(
\begin{array}{cccc}
  \rho_{00} & 0 & 0 & \rho_{03} \\
  0 & \rho_{11} & \rho_{12} & 0 \\
  0 & \rho_{21} & \rho_{22} & 0 \\
  \rho_{30} & 0 & 0 & \rho_{33} \\
\end{array}\right).
\label{dmatrix}
\end{equation}
For this $X$ state, the measurement on subsystem $B$ leads to ensembles $\{\rho_0,p_0\}$ and $\{\rho_1,p_1\}$
such that the quantum conditional entropy becomes:
\begin{equation}
\emph{S}(\rho_{AB}|\{\Pi_{x}\})=p_{0}S(\rho_{0})+p_{1}S(\rho_{1}),
\label{min}
\end{equation}
where $\rho_{0}$, $\rho_{1}$ are density matrices of the subsystem $A$ after measuring the subsystem $B$, and $p_{0}$, $p_{1}$ are their respective probabilities.
The corresponding eigenvalues are $\mathit{v}_\pm(\rho_0)=(1\pm\theta)/2$ and $\mathit{w}_\pm(\rho_1)=(1\pm\theta^\prime)/2$ where $\theta$, $\theta^\prime$ are defined in ref.~\cite{xstate}. The optimization has been reduced to a four parameter problem with four possible solutions which are extremes of the functional  (\ref{min}) \cite{xstate}. We shall look for the set of parameters among these solutions which lead to the minimum value of the quantum conditional entropy.

In what follows we consider two specific physical scenarios where the state (\ref{gstate}) can be physically realized.
Following the recipes outlined in previous discussion we are able to calculate EoF analytically as a function of time for the mixed state of the $(2\otimes d)$ party.

Our first example considers two atoms ($A$ and $B$), each of them interacting resonantly with a single quantized mode of a cavity field (system $C$) in a Fock state.
This physical situation is described through the two-atom Tavis-Cummings (TC) Hamiltonian:
$H=\hbar g[(\sigma_{A}+\sigma_{B}) a^{\dagger}_C+(\sigma_{A}^{\dagger}+\sigma_{B}^{\dagger})a_C]$,
where $\sigma_j$ and $\sigma_{j}^{\dagger}$ are the ladder Pauli operators for the $j$\emph{th} atom, $a$ ($a^{\dagger}$)
is the annihilation (creation) operator for photons in cavity $C$ and $g$ is the coupling constant.
Let us assume that the system is initially in the state $|\Psi(0)\rangle=(\alpha|gg\rangle+\beta|ee\rangle)_{AB}|n\rangle_C$.
Since the TC Hamiltonian preserves the total number of excitation, the cavity mode will evolve within a $5$-dimensional Hilbert space spanned by $\{|n-2\rangle,|n-1\rangle,|n\rangle,|n+1\rangle,|n+2\rangle\}$ for $n\ge 2$.
For $n=0, 1$ the dimension will be $3$ and $4$ respectively.
On the other hand, the atomic system will evolve within the subspace $\{|gg\rangle,|+\rangle,|ee\rangle\}$ with $|+\rangle=(|eg\rangle+|ge\rangle)/\sqrt 2$.
By Solving the Schr\"{o}dinger equation, the system at time $t$ is described by the state:
\begin{eqnarray}
|\Psi(t)\rangle&=&c_{1}(t)|gg\rangle|n+2\rangle+c_{2}(t)|+\rangle|n+1\rangle \notag \\
&+&c_{3}(t)|ee\rangle|n\rangle+c_{4}(t)|gg\rangle|n\rangle \notag \\
&+&c_{5}(t)|+\rangle|n-1\rangle+c_{6}(t)|ee\rangle|n-2\rangle,
\label{solutionn}
\end{eqnarray}
where the probability amplitudes are
\begin{small}
\begin{subequations}\label{probamn}
\begin{eqnarray}
c_{1}(t)\!&=&\!-\frac{\beta\sqrt{(n\!+\!1)(n\!+\!2)}}{2n\!+\!3}\left[1\!-\!\cos(\sqrt{2(2n\!+\!3)}gt)\right],  \label{c1} \\
c_{2}(t)\!&=&\!-\frac{i\beta\sqrt{n\!+\!1}}{\sqrt{2n\!+\!3}}\sin(\sqrt{2(2n\!+\!3)}gt),  \label{c2} \\
c_{3}(t)\!&=&\!\beta\left\{1-\frac{(n\!+\!1)}{2n\!+\!3}\left[1\!-\!\cos(\sqrt{2(2n\!+\!3)}gt)\right]\right\},  \label{c3} \\
c_{4}(t)\!&=&\!\alpha\left\{1-\frac{n}{2n\!-\!1}\left[1-\cos(\sqrt{2(2n\!-\!1)}gt)\right]\right\},  \label{c4}\\
c_{5}(t)\!&=&\!-\frac{i\alpha\sqrt{n}}{\sqrt{2n\!-\!1}}\sin(\sqrt{2(2n\!-\!1)}gt),  \label{c5} \\
c_{6}(t)\!&=&\!-\frac{\alpha\sqrt{n(n\!-\!1)}}{2n\!-\!1}\left[1\!-\!\cos(\sqrt{2(2n\!-\!1)}gt)\right].  \label{c6}
\end{eqnarray}
\end{subequations}
\end{small}
The entanglement of formation $E_{AC}$ between the atom $A$ and the cavity mode $C$ is given in terms of the quantum correlations embedded in the $AB$ subsystem.
The reduced density matrix $\rho_{AB}$ has the form of an $X$ state~(\ref{dmatrix}). The matrix elements in that representation are
$\rho_{00}=|c_1|^2+|c_4|^2$, $\rho_{11}=\rho_{22}=\rho_{12}=\rho_{21}=(|c_2|^2+|c_5|^2)/2$, $\rho_{33}=|c_3|^2+|c_6|^2$, and $\rho_{03}=\rho_{30}^*=c_3 c_4^* $.
$E_{AC}$ is the minimum of $S(\rho_{AB}\mid\{\mid\Pi_x\})$ in equation (\ref{min}) and it is obtained analytically through the optimal eigenvalues
$\{v_\pm (\theta),w_\pm(\theta^\prime)\}$ of the density matrices $\rho_0$ and $\rho_1$ \cite{xstate}.
\begin{figure}[t]
\includegraphics[width=65mm]{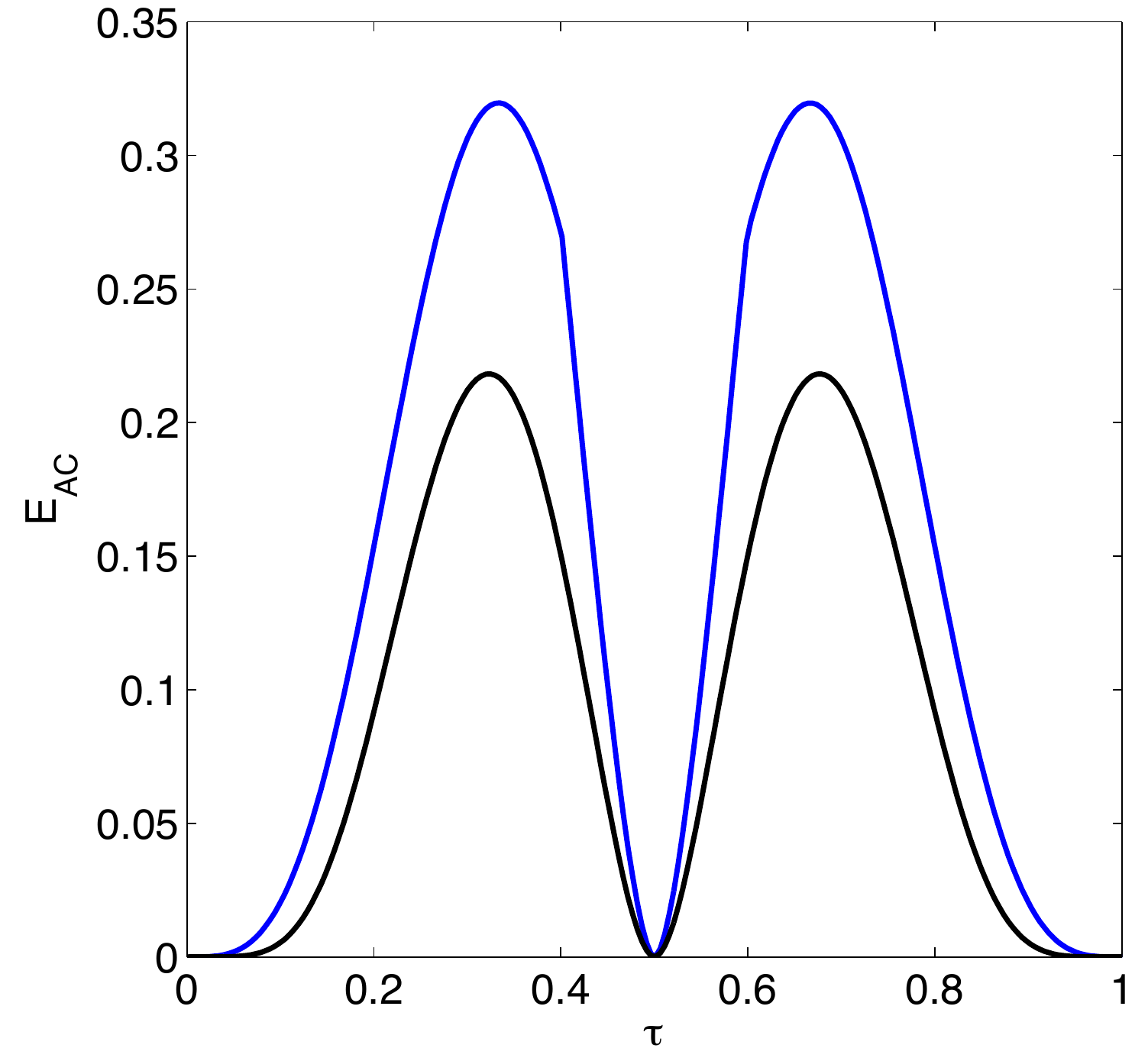}
\caption{Entanglement of formation between the atom $A$ and the cavity mode $C$ for the initial  state
$|\Psi(0)\rangle=(\alpha|gg\rangle+\beta|ee\rangle)_{AB}|0\rangle_C$  with $\alpha=\beta=1/\sqrt{2}$. Blue line corresponds to analytical expression and black line to lower bound of EoF as a function of the dimensionless time $\tau=\sqrt{6}gt/2\pi$.} \label{fig1}
\end{figure}

As an example, let us consider the case $n=0$ where the total system has dimension $2\otimes2\otimes3$.
In such case we found that there are two sets of parameters that can  minimize Eq.~(\ref{min}), with corresponding eigenvalues:
$\{v_\pm (\theta_1)=w_\pm (\theta^\prime_1)\}$ and $\{v_\pm(\theta_2),w_\pm(\theta^\prime_2)\}$, where
\begin{small}
\begin{eqnarray}
\theta_1\!&=&\!\theta^\prime_1\!=\!\sqrt{(\alpha^{2}+c_1^2-c_3^2)^{2}+4(\alpha c_3 + \frac{1}{2}|c_2 |^2)^{2}}, \\
\theta_2\!&=&\!\frac{\left|\alpha^2+c_1^2 -\frac{1}{2}|c_2 |^2\right|}{\alpha^2+c_1^2 +\frac{1}{2}|c_2 |^2}, \quad \theta^\prime_2\!=\!\frac{\left|c_3^2 -\frac{1}{2}|c_2 |^2\right|}{c_3^2 +\frac{1}{2}|c_2 |^2}.
\end{eqnarray}
\end{small}
At a time $t$ the EoF is
\begin{equation}
E_{AC}=\min\{S(\theta_1,\theta^\prime_1),S(\theta_2,\theta^\prime_2)\}, \label{case1}
\end{equation}
where we have written $S(\theta,\theta')$ for the conditional entropy (\ref{min}).
For $S(\theta_1,\theta^\prime_1)$, $p_0=p_1=1/2$ and for $S(\theta_2,\theta^\prime_2)$, $p_{0}=\alpha^2+c_1^2 +|c_2|^2/2$ and $p_{1}=1-p_0$.
The dynamics of $E_{AC}$ is shown in Fig.~\ref{fig1} where we restricted the dimensionless time $\tau=\sqrt{6}gt/2\pi$ to one period.
We observe that the entanglement is zero for $\tau=1/2$ where the $(2\otimes3)$-dimensional  separable state given by $\rho=|g\rangle \langle g | \otimes |\phi\rangle\langle\phi|+\beta^2/9|e\rangle \langle e| \otimes |0\rangle\langle 0|$,
with $|\phi\rangle=\alpha|0\rangle-2\beta\sqrt{2}/3|2\rangle$.
Although there is no entanglement, this state is maintaining quantum correlations as can be checked by calculating the quantum discord which is different from zero except for $\alpha =0$ (classical state) and $\alpha =1$ (factorized state).
The amount of discord in such case is called dissonance \cite{Modi}. For $\tau=1$ the system is again in the factorized initial state. From Eqs.~(\ref{probamn}) we realize that the arising of entanglement is related with the population $c_2(t)$ of the symmetric state $|+\rangle$, e.i., entanglement will arise whenever $c_2(t)\neq 0$.

From Fig.~\ref{fig1} we see that the entanglement dynamics suffers \emph{abrupt changes}. These happend
for dimensionless times $\tau\approx0.40098$ and $\tau\approx 0.59902$. These behavior arise because there are two solutions leading to equation (\ref{case1}).
For $\tau\lesssim0.40098$ the minimum conditional entropy is $S(\theta_2,\theta_2^\prime)$.
For $0.40098\lesssim\tau\lesssim0.59902$ the minimum conditional entropy is  $S(\theta_1,\theta_1^\prime)$.
For $\tau\gtrsim 0.59902 $ the solution is again  $S(\theta_2,\theta_2^\prime)$. These \emph{abrupt changes} appear for all $\alpha$ different to zero.
Evidence for the existence of \emph{abrupt changes} in a lower bound for disentanglement dynamics in higher dimensional systems has been previously reported \cite{las}. 
\begin{figure}[t]
\includegraphics[width=63mm]{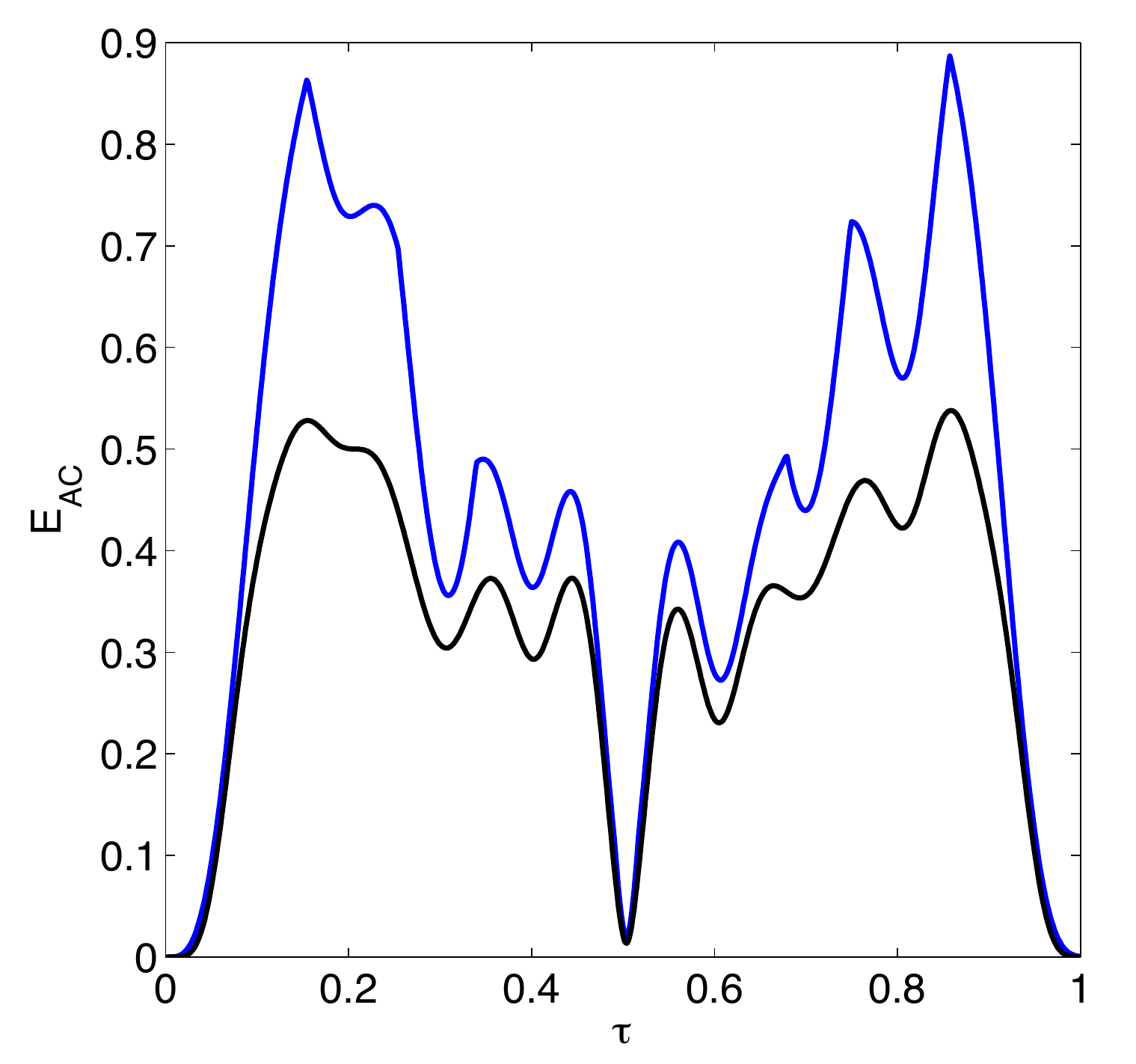}
\caption{Entanglement of formation (blue line) and lower bound (black line)  for a $(2\otimes 5)$-dimensional  state. The  initial state is
$(\alpha|gg\rangle_{AB}+\beta|ee\rangle_{AB})\otimes |2\rangle_C$ with  $\alpha=\beta=1/\sqrt{2}$ and the dimensionless time $\tau=\sqrt{14} g t/ 6\pi$.} \label{fig2-1}
\end{figure}

By assuming the initial state $|\Psi(0)\rangle=(\alpha|gg\rangle+\beta|ee\rangle)_{AB}|n\rangle_C$
the atom-field system will evolve inside a $(2\otimes4)$-dimensional Hilbert space when $n=1$ and inside a $(2\otimes5)$-dimensional Hilbert space when $n\ge2$.
Fig.~\ref{fig2-1} shows the evolution of the $E_{AC}$ for $n=2$. In this case the symmetric state associated to the cavity state $|n-1\rangle$ is also populated, i.e.,
entanglement arises when the probability amplitudes $c_2(t)$ and $c_5(t)$ are different from zero.
We notice that at $\tau=1/2$ the entanglement approaches to zero.
At that dimensionless time the atom $A$ and the field mode $C$ are near the factorized state
$\rho=|g\rangle\langle g|\otimes|\phi\rangle\langle\phi|$ with $|\phi\rangle=\beta|2\rangle+\alpha|4\rangle$.
In other words, at $\tau=1/2$ the state of the atomic subsystem has been approximately swapped to the field state.
The minimum observed at $\tau=1$ corresponds approximately to the initial state.
On the other hand, for this initial state we found that there are three set of parameters that can minimize Eq. (\ref{min}), that is,
the EoF will corresponds to $E_{AC}=\min\{S(\theta_1,\theta^\prime_1),S(\theta_2,\theta^\prime_2),S(\theta_3,\theta^\prime_3)\}$
where the entropies can be obtained following the previous procedure.
The dependence on three different entropies is the responsible for the \emph{abrupt changes} showed in Fig.~\ref{fig2-1}.

Figs. \ref{fig1} and \ref{fig2-1} shows the lower bound \cite{Chen05} used to compare with EoF. In both cases it follows monotonically the EoF, however it does not exhibit the abrupt changes.

Different initial conditions for this system can be considered whose evolution result in a $X$-type state for the atomic subsystem.
In particular, initial entanglement between one atom and the cavity mode can be considered.
For instance, $|\Psi(0)\rangle=\alpha|gg\rangle|n+1\rangle+\beta|ge\rangle|n\rangle$ and $|\Psi(0)\rangle=\alpha|eg\rangle|n+1\rangle+\beta|ee\rangle|n\rangle$. 

A second physical scenario is a system consisting of two atoms $A$ and $B$ interacting with a common reservoir $C$.
After tracing out the reservoir degrees of freedom the dynamics of the atomic subsystem is described by the master equation:
$\dot\rho=(\gamma / 2)( 2 J \rho J^{\dagger}-J^{\dagger }J \rho -\rho J^{\dagger}J)$, with $J=\sigma_{A}+\sigma_{B}$.
This model has been used to describe the dissipative dynamics of a collection of cold ions coupled through the center-of-mass mode \cite{milburn02}. However, by 
considering explicitly the reservoir degrees of freedom and initially in the vacuum state the overall system will evolve in a $(2\otimes2\otimes3)$-dimensional Hilbert space.
\begin{figure}[t]
\includegraphics[width=65mm]{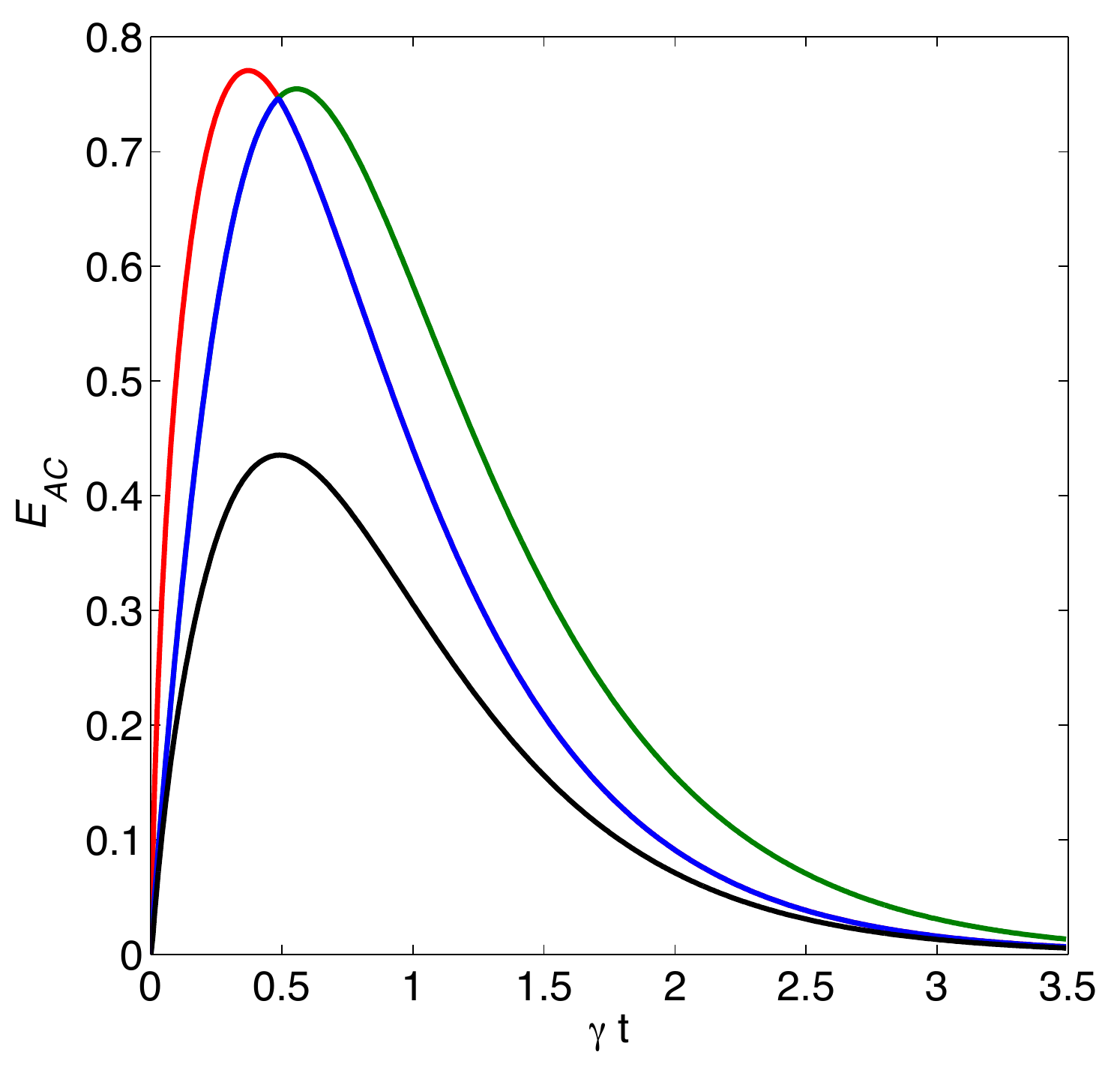}
\caption{Entanglement of formation between atom $A$ and cavity mode (blue line) for the initial state
$|\Psi(0)\rangle=(\alpha|gg\rangle+\beta|ee\rangle)_{AB}|\bar{0}\rangle_C$ with $\alpha=0.3$. The red and green lines shows the evolution of the conditional
entropies $S(\theta_1,\theta_1^\prime)$ and $S(\theta_2,\theta_2^\prime)$ respectively. Black line corresponds to the lower bound.}
\label{fig3}
\end{figure}
Let us assume the initial state
$|\Psi(0)\rangle=(\alpha|gg\rangle+\beta|ee\rangle)_{AB}|\bar{0}\rangle_C,$ where $|\bar{0}\rangle=\prod_k |0\rangle_k$ is the reservoir  the vacuum state.
Inspecting the occupied Hilbert space throughout the evolution we found that the overall state is given by:
\begin{eqnarray}
|\Psi(t)\rangle &=& \alpha|gg\rangle_{AB}|\bar{0}\rangle_{C}+c_{1}(t)|ee\rangle_{AB}|\bar{0}\rangle_{C}  \notag \\
&+&c_{2}(t)|+\rangle_{AB}|\bar{1}\rangle_{C}+c_{3}(t)|gg\rangle_{AB}|\bar{2}\rangle_{C},
\label{solution3}
\end{eqnarray}
where $|\bar{k}\rangle$  are collective states of the  reservoir having $k$ excitations \cite{lop}. In this case the probability amplitudes become $
c_{1}(t)=\beta e^{-\gamma t}$, $c_{2}(t)=\beta e^{-\gamma t}\sqrt{1-e^{-2\gamma t}}$ and $c_{3}(t) = \sqrt{1-\alpha^{2}-c_{1}^{2}(t)-c_{2}^{2}(t)}$.

This evolution takes to the atomic subsystem to be an $X$ state where the matrix elements
$\rho_{00}=c_{1}^{2}$, $\rho_{03}=\alpha c_{1}$, $\rho_{11} = \rho_{22}=\rho_{12}=\rho_{21}=c_{2}^{2}/2 $ and $\rho_{33}=\alpha^{2}+c_{3}^{2}$.
Here, we have also two relevant sets of parameters minimizing Eq.~(\ref{min}), leading to eigenvalues $\{v_\pm (\theta_1) = w_\pm (\theta^\prime_1)\}$ and
$\{v_\pm (\theta_2),w_\pm (\theta^\prime_2)\}$, with
\begin{eqnarray}
\theta_1 & = & \theta^\prime_1=e^{-4 \gamma t}[\beta ^2 (2 e^{3 \gamma t} \alpha +e^{2 \gamma t} \beta -\beta ^2)^2
\notag \\
&&\qquad+(e^{4 \gamma t}-3 e^{2 \gamma t} \beta ^2+\beta ^2)^2]^{1/2}, \notag\\
\theta_2 & = & \frac{1 +e^{2 \gamma t}}{3 e^{2 \gamma t}-1}, \quad
\theta^\prime_2 = \frac{\left|2 e^{4 \gamma t}-5 e^{2 \gamma t} \beta ^2+3 \beta ^2\right|}{\left|2 e^{4 \gamma t}-3 e^{2 \gamma t} \beta ^2+\beta ^2\right|}.  \notag
\end{eqnarray}
In this case, the EoF is again given by Eq. (\ref{case1}) with $p_{0}=e^{-4\gamma t}(3e^{2\gamma t}-1)\beta^2 /2$ for $S(\theta_2,\theta^\prime_2)$
and $p_0=p_1=1/2$ for $S(\theta_1,\theta^\prime_1)$.
Fig.~\ref{fig3} shows the two relevant conditional entropies $S(\theta_1,\theta^\prime_1)$ and $S(\theta_2,\theta^\prime_2)$, the EoF, and a lower bound. In this case the dynamics between atom $A$ and the reservoir evolves in a ($2\otimes 3$) Hilbert space.
We observe that there is only one \emph{abrupt change} along the entanglement dynamics and that the atom-reservoir state converge to the factorized state
$|gg\rangle(\alpha|\bar{0}\rangle+\sqrt{1-\alpha^2}|\bar{2}\rangle)$ for time $t>>1/\gamma$. Notice that the reservoir converge to an coherent superposition when $\alpha\neq0$.
The \emph{abrupt change} holds at time for which the conditional entropies are equal and it depends on the initial probability amplitudes. As in previous cases, the lower bound has a monotonic behavior with respect to the EoF and  it does not exhibit the \emph{abrupt change}.

In summary, we have obtained an analytical recipe to evaluate  the entanglement of formation for a class of $2\otimes d$ bipartite mixed states. For this class of states the calculation of EoF is mapped to an optimization process for calculating a conditional entropy in a $2\otimes 2$ partition. Previous result for lower bounds shows to be in agreement with our findings. Finally, we have found \emph{abrupt changes} in the entanglement dynamics  as consequence of the optimization process leading to the minimum value of the quantum conditional entropy. 

These results open the possibility for extending this treatment to evaluate EoF for arbitrary dimensions aided by such optimization for conditional entropies which can be realized numerically at least.

\begin{acknowledgments}
C.E.L. acknowledges financial support from DICYT 041131LC and PBCT-CONICYT PSD54, L.R. from Fondecyt 1080535 and J.C.R. from Fondecyt 1100700.
F.L, C.E.L. and J.C.R. acknowledge support from Financiamiento Basal para Centros Cient\'{\i}ficos y Tecnol\'ogicos de Excelencia.
\end{acknowledgments}

\end{document}